\documentstyle{article}

\begin{document}

\title{A systematic treatment of triaxial systems at high spins. II}
\author{J. P. Garrahan\thanks{Fellow of the CONICET,
Buenos Aires, Argentina. E-mail: garrahan@tandar.edu.ar} and D. R.
Bes\thanks{Fellow of the CONICET, Buenos Aires,
Argentina.  E-mail:  bes@cnea.edu.ar} \\ Departamento de F\'{\i}sica,
C.N.E.A., Av. del Libertador 8250, (1429) \\ Buenos Aires, Argentina}
\date{October 15, 1993}

\maketitle

\begin{abstract}
The BRST treatment of triaxial nuclei rotating at high spins, which had
been previously developed, is extended  to relate states with
neighboring values of the angular momentum $I$. As a consequence, the
irreducible multipole transition operators can be unambiguously
separated into their physical and spurious components.  As a by
product, we give explicit expressions for the operators associated with
the Goldstone-Nambu modes in terms of the final orthonormal set of
physical and spurious modes.
\end{abstract}

\section{Introduction} \label{sint}
The ellipsoidal rigid top is the simplest model which can be used to
describe the properties of nuclei rotating at high spin. It includes
the ``vibrational'' wobbling motion \cite{BM75}. The next refinement in
the treatment of such systems is the cranking model \cite{Go76,BF79},
which has become the indispensable tool for the evaluation of many
nuclear properties. It is particularly successful in the prediction of
energy levels, single-particle states, deformation parameters,
etc\ldots

The problem of constructing the normal modes associated with the
cranking model has been treated by several authors
\cite{Ma77,MJ78,Ma79,SM83}.  All these treatments assume the quadrupole
plus (monopole) pairing model.  They are made either in the uniformly
rotating frame or in the principal-axis frame.  The collective degrees
of freedom do not appear explicitly in these treatments.  As a
consequence, there are no properly spurious modes, but rather redundant
modes which, for instance, represent the system with a different
orientation \cite{Ma79}.  If the angular momentum is oriented along the
$x$-axis, there appear wobbling modes in the perpendicular directions,
as many as solutions of the RPA equations  which are not of the
Goldstone-Nambu type.

A different treatment of the cranking model at high spins has been
presented in ref. \cite{KBC90}. It is based on the application of the
BRST symmetry \cite{BRST,HT92} to many-body problems \cite{BK90}. Our
gauge condition requires the vanishing of the RPA angles, which are
obtained in the usual way \cite{MW70} for the $x$-direction (cf. eq.
(\ref{a122})) or through a suitable generalization for the
perpendicular directions (cf. eqs. (\ref{a777})). The collective
coordinates are explicitly introduced. Consequently, there is a
wobbling mode which appears together with the GN modes of the RPA
equations.  The treatment of ref.  \cite{KBC90} is able to separate
this physical mode from the GN modes, which are incorporated to the
spurious sector.

In the BRST treatment there are no restrictions concerning the
hamiltonian, other than those associated with the validity of the
cranking model (small fluctuations in the collective angular momentum
$\vec{I}$ relative to the expectation value $\langle I_x \rangle$, and
small fluctuations in the deformations relative to the equilibrium
values).  In particular, the velocity-dependent residual interactions,
insuring the Galilean invariance of the hamiltonian, may be treated on
the same footing as the interactions which are directly responsible for
the mean field. The importance of terms of this type has been recently
stressed in ref. \cite{HN93}.

The Bohr collective hamiltonian at high spins is another example to
which the methods developed for quadrupole + pairing hamiltonians
cannot be directly applied. In such case we have performed perturbative
corrections to the lowest order cranking results \cite{GB93}.

In the present paper, we review the results of ref. \cite{KBC90} in
subsection \ref{ssbrst} and in appendix A. We explicitly give here the
inverted form of the equations yielding the normal modes in terms of
the set of variables associated with the GN modes and with the
collective rotations (appendix A). This step is necessary for
constructing the vertices of the residual interaction in perturbative
calculations which may be eventually performed.

The main object of the paper is to construct the matrix elements of the
transition operators within the formalism. In the first place, we treat
states associated with neighboring values of the angular momentum $I$
(subsection \ref{iota}). Subsequently,  we study the matrix elements of
multipole operators between the BRST basic set of states (section
\ref{sirr}). Intensive use is made of the algebra developed by
Marshalek to treat the collective coordinates \cite{Ma75}, which is
reviewed in appendix B.

The present work may be considered to be the continuation of ref.
\cite{KBC90}. Hence the label II in the title.

\section{The BRST procedure} \label{sbrst}
\subsection{Review of the formalism} \label{ssbrst}
A  suitable procedure to treat perturbatively systems displaying both
collective and intrinsic degrees of freedom is developed in ref.
\cite{BK90} as an application of the BRST symmetry. It is based on the
fact that a mechanical system which is described from a moving frame of
reference is a gauge system in (0+1) dimensions, the gauge invariance
arising from the physical equivalence between moving the system in one
direction and the frame in the opposite one.  To choose a gauge is to
fix a relation between the moving frame and the system. The constraints
are obtained from the non-inversibility of some the equations
expressing the momenta in terms of the velocities.

For systems with triaxial deformations, the collective degrees of
freedom are expressed through the three Euler angles describing the
orientation of the moving frame, and by the three collective angular
momentum components $I_{\nu}$ in the moving frame (eq. (\ref{ies})).
Since the collective variables are raised to the level of true
variables, some combination of the intrinsic variables (i.e., those
associated with each nucleon) belong to the spurious sector. This
sector is completed through the introduction of three Lagrange
multipliers $\Omega_{\nu}$ (and their conjugate momenta $P_{\nu}$) and
six ghost (fermion) variables $\eta_{\nu},{\bar \eta}_{\nu}$ (and their
conjugate partners $\pi_{\nu},{\bar \pi}_{\nu}$).

The six constraints
\begin{equation}
f_{\nu}\equiv J_{\nu}-I_{\nu}=0 \; ; \;\;\;\;\; P_{\nu}=0, \label{x1}
\end{equation}
where $J_{\nu}$ are the intrinsic angular momentum operators
($[J_{\mu},J_{\nu}]=i\epsilon_{\mu \nu \rho}J_{\rho}$), define the
(nilpotent and hermitian) BRST ``charge''
\begin{equation}
{\cal Q} = {\bar \pi}_{\nu} P_{\nu} - \eta_{\nu} (J_{\nu}-I_{\nu}) +
		\half \mathrm{i} \epsilon_{\mu \nu \rho}
		\eta_{\mu}  \eta_{\nu} \pi_{\rho} \label{x2}
\end{equation}

Within the BRST formalism the constraints (\ref{x1}) are replaced by
the requirement to act within the subspace of states that are
annihilated by the BRST charge. This subspace (henceforth, the physical
subspace)  consists of states which are annihilated by the constraints
(\ref{x1}), plus zero-norm states $|\mathrm{null}\rangle$.
Consistently, the enlarged physical operators are defined as those
commuting with the charge ${\cal Q}$.  They include the operators
commuting with the constraints (\ref{x1}) plus the ``null'' operators
$O_\mathrm{null}$, which transform a physical state into a
$|\mathrm{null}\rangle$ state
\begin{equation}
|\mathrm{null}\rangle \equiv {\cal Q}|u\rangle \; ; \;\;\;\;\;
	O_\mathrm{null} \equiv [O_\mathrm{u},{\cal Q}]_{\pm}
\end{equation}
Such enlargement of the concept of physical states and operators allows
the necessary freedom in order to solve the constrained problem through
an unconstrained formalism.

The BRST hamiltonian, with a suitable choice of $H_\mathrm{u}$, reads
\begin{eqnarray}
H_{\mathrm{BRST}} &=& H + [H_\mathrm{u},{\cal Q}]_+ \nonumber \\
	 &=& H - \Omega_{\nu} (J_{\nu} - I_{\nu}) + \frac{1}{F_{\nu}}G_{\nu}P_{\nu} -
 		\frac{A_{\nu}}{2 F_{\nu}^2} P_{\nu}^2 \nonumber\\
	 	& & + \mathrm{i} \pi_{\nu} {\bar \pi}_{\nu} -
		\frac{1}{F_{\nu}}  {\bar \eta}_{\mu}
		\eta_{\nu}  [G_{\mu},J_{\nu}] +
  		\mathrm{i}  \epsilon_{\mu \nu \rho}
		\Omega_{\rho} \pi_{\mu} \pi_{\nu} \label{x3}
\end{eqnarray}

Three of the six arbitrary parameters $A_{\nu},F_{\nu}$ are eliminated at
subsequent steps of the formalism (eqs. (\ref{ax}) and (\ref{ayz})). The
remaining three give rise to the frequencies of the excitations in the
spurious sector (cf. eq. (\ref{z5})) and should disappear from any
physical result.

The hamiltonian $H_{BRST}$ is minimized under the conditions
\begin{eqnarray}
\langle I_x \rangle = I \; ; \;\;\;\;\;
		\langle I_y \rangle =
		\langle I_z \rangle =0    \label{z2}
\end{eqnarray}
As a result of the minimization we obtain the expectation values
\begin{eqnarray}
\langle P_{\nu} \rangle &=& \langle \Omega_y \rangle =
		\langle \Omega_z \rangle = 0
		\nonumber \\
		\langle \Omega_x \rangle &=& I/\Im \label{z3}
\end{eqnarray}
The particle basis and the moment of inertia $\Im(I)$ are determined
from the minimization of the routhian and the self-consistency
condition
\begin{eqnarray}
\delta \langle (H- \langle \Omega_x \rangle J_x) \rangle = 0
		\; ; \;\;\;\;\;
		\langle J_x \rangle = I \label{z4}
\end{eqnarray}

In the next step, the normal modes associated with the BRST hamiltonian
are obtained. As usual in cranking calculations, the modes can be
classified according to their behaviour relative to the rotation over
an angle $\pi$ about the $x$-axis \cite{Go76}. The corresponding
quadratic hamiltonians\footnote{In boson systems, the superscripts
$(\nu)$ denote terms with $\nu$ creation and annihilation operators. In
fermion systems, the superscript {(1)} indicates terms which are linear
in the RPA quasi-bosons; {(2)} means those which become quadratic upon
bosonization (such as the particle-particle terms in one-body
operators), etc.} read
\begin{eqnarray}
H_x^{(2)} &=& \omega_{rx}(\Gamma^{\dag}_{rx}\Gamma_{rx}+\half) +
	\omega_x(\Gamma^{\dag}_{1x}\Gamma_{1x} -
	\Gamma^{\dag}_{0x}\Gamma_{0x}+{\bar a}_xa_x+{\bar b}_xb_x)
	\nonumber \\
H_{\perp}^{(2)} &=& \omega_{rp}
		(\Gamma^{\dag}_{rp}\Gamma_{rp} + \half) +
		\omega_\mathrm{w}
		(\Gamma^{\dag}_{\mathrm{w}} \Gamma_{\mathrm{w}} + \half)
		\nonumber \\
		& & + \omega_p
		(\Gamma^{\dag}_{1p}\Gamma_{1p} -
		\Gamma^{\dag}_{0p}\Gamma_{0p} +
		{\bar a}_p a_p+{\bar b}_p b_p) \label{z5}
\end{eqnarray}
Here $\omega_{rv}$ ($v=x,p$; $p=\pm$) are the RPA eigenfrequencies of
the finite, real modes. The wobbling mode has been disentangled from
the spurious sector, as shown in \cite{KBC90} and summarized in appendix
A.  Its frequency is
\begin{equation}
\omega_\mathrm{w} = I \Delta_y \Delta_z  \; ; \;\;\;\;\;
		\Delta_{y/z}=
		\sqrt{\frac{1}{\Im_{y/z}}-\frac{1}{\Im}} \label{z80}
\end{equation}
where the moments of inertia $\Im_{y/z}$ are obtained from eqs.
(\ref{a777}). The frequencies $\omega_x, \omega_p$ are arbitrary
parameters associated with the spurious sector, which is clearly
separated from the physical one.  The creation and annihilation
operators of the spurious sector satisfy the following non-vanishing
commutation and anticommutation relations
\begin{eqnarray}
[\Gamma_{1 \nu},\Gamma^{\dag}_{1 \nu}] =
		[\Gamma^{\dag}_{0 \nu},\Gamma_{0 \nu}]=
		[{\bar a}_{\nu},a_{\nu}]_+ =
		[{\bar b}_{\nu},b_{\nu}]_+= 1 \label{z6}
\end{eqnarray}

The quadratic hamiltonians (\ref{z5}) determine both the real and the
spurious spectra associated with a given value of $I$. The
corresponding states span a space which is factorized into real,
spurious and rotational subspaces
\begin{eqnarray}
|n_{r \nu},n_{\mathrm{w}}\rangle_I \,
		|n_{0 \nu},n_{1 \nu},n_{a \nu},n_{b \nu}\rangle_I \,
		|IM\rangle =
		{\cal N} \, \prod_{\nu}
		(\Gamma^{\dag}_{r \nu})^{n_{r \nu}} \,
		(\Gamma^{\dag}_{\mathrm{w}})^{n_{\mathrm{w}}} \,
		|\mathrm{r}\rangle_I
		\nonumber \\
		\times
		(\Gamma^{\dag}_{0 \nu})^{n_{0 \nu}} \,
		(\Gamma^{\dag}_{1 \nu})^{n_{1 \nu}} \,
		({\bar a_{\nu}})^{n_{a \nu}}
		({\bar b_{\nu}})^{n_{b \nu}} \,
		|\mathrm{sp}\rangle_I \,
		|IM\rangle \label{z7}
\end{eqnarray}
where  $n_{r \nu}, n_\mathrm{w},n_{1 \nu},n_{0 \nu}=0,1,2,...$ and
$n_{a \nu},n_{b \nu}=0,1$.  In particular, the spurious sector is
labelled by the quantum numbers $n_{0 \nu},n_{1 \nu},n_{a \nu},n_{b
\nu}$. It includes: i) unphysical states (i.e., states which are not
annihilated by the BRST charge); ii) zero-norm physical states (for
instance, states with $n_{0 \nu}=n_{1 \nu}$); iii) a single
normalizable physical state, namely the vacuum state
$|\mathrm{sp}\rangle_I$.  The product of the vacua of the real and
spurious excitations ($|\mathrm{r}\rangle_I$ and
$|\mathrm{sp}\rangle_I$, respectively) is also the yrast state for a
given $I$. The rotational states $| IM \rangle$ are, in configuration
space, the wave functions $D_{MI}^I$ of the rigid top.  The rotational
excitations $D^I_{MK}$ ($K<I$) are partly associated with the wobbling
motion and partly with the spurious sector. Here $K$ is the projection
along the intrinsic $x$-axis.

\subsection{Excitations with $I'= I+\iota$ }  \label{iota}
Since our aim is to extend the formalism to include transition
probabilities, we must perform two tasks:

\begin{enumerate}
\item[(i)] to look for modifications in the residual hamiltonian when
considering a state with $I'=I+\iota$, within the basis (\ref{z7})
associated with the angular momentum $I$.

The additional terms in the residual hamiltonian are due to changes in
the parameters entering in the BRST term\footnote{Throughout the paper
we use the spherical representation $O_{\sigma} = -
\frac{\sigma}{\sqrt{2}}(O_y +i\sigma O_z)$, $\sigma = \pm 1$}
\begin{eqnarray}
-\Omega_{\nu} (J_{\nu}-I_{\nu})|_{I+\iota} &\rightarrow&
		-\frac{I}{\Im}J_x + \Omega_x I -
		\Omega_{\sigma}I_{-\sigma} \nonumber \\
		& & +\iota
		\left[ \left(1-\frac{I}{\Im}\frac{\d \Im}{\d I} \right)
		\frac{J_x}{\Im} +\Omega_x^{(1)} -
		\frac{\Omega^{(1)}_{\sigma}I^{(1)}_{-\sigma}}{2I} +
		{\cal O}(I^{-\frac{3}{2}}) \right] \label{z8}
\end{eqnarray}
Using the fact that $J^{(1)}_x$ is a null operator and the expression
for $\Omega_x^{(1)}$ (eq. (\ref{diagx})) the linear term in the
second line of eq. (\ref{z8}) can be written
\begin{equation}
H^{(1)}_{\iota} = -\iota \sqrt{\frac{\omega_x}{2\Im_x}} \,
		(\Gamma^{\dag}_{1x}+\Gamma_{1x}) +
		O_\mathrm{null} \label{z10}
\end{equation}
which corrects the vacuum of the spurious sector
\begin{eqnarray}
|\mathrm{sp}\rangle_{I+\iota} &=&
		(1 + \iota  \frac{1}{\sqrt{2 \omega_x \Im_x}}
		\Gamma^{\dag}_{1x}) \, |\mathrm{sp}\rangle_I +
		|\mathrm{null}\rangle \nonumber\\
		&=& (1 + \mathrm{i} \iota \theta_x) \,
		|\mathrm{sp}\rangle_I +
		|\mathrm{null}\rangle,   \label{z11}
\end{eqnarray}
according to eq. (\ref{diagx}).  Therefore the spurious ground state
associated with the value $I+\iota$ of the angular momentum adquires an
unphysical component if expressed in terms if the states corresponding
to the angular momentum $I$.

We should verify that the states (\ref{z11}) are annihilated by the
modified BRST charge (\ref{x2}). To leading order,
\begin{eqnarray}
{\cal Q}_{\iota} &=& {\cal Q} + \iota \eta_x \nonumber \\
{\cal Q}_{\iota}|\mathrm{sp}\rangle_{I+\iota} &=&
		\iota (\eta_x+ \mathrm{i} [{\cal Q},\theta_x]) \,
		|\mathrm{sp}\rangle_I = 0 \label{z111}
\end{eqnarray}

\item[(ii)] to determine the form of the operators connecting states
corresponding to different values of $I$. This is accomplished in
subsection \ref{del}.
\end{enumerate}

\section{Irreducible transition operators}  \label{sirr}
\subsection{The physical nature of the transition operators}
\label{sphys}
The irreducible transition operators in the laboratory and in
the moving frames are related by
\begin{equation}
T^{\mathrm{(lab)}}_{\lambda \mu} =
		D^{\lambda}_{\mu \nu} T_{\lambda \nu}, \label{y1}
\end{equation}
the tensor properties being expressed by the relations
\begin{eqnarray}
{[J_{\sigma},T_{\lambda \nu}]}= -\sigma \,
	\sqrt{(\lambda - \sigma \nu)(\lambda + \sigma \nu +1)/2} \,
	T_{\lambda \nu + \sigma} \; ; \;\;\;\;\;
	[J_x,T_{\lambda \nu}]=
	\nu T_{\lambda \nu} \nonumber \\
{[I_{\sigma},D^{\lambda}_{\mu \nu}]}= -\sigma \,
	\sqrt{(\lambda + \sigma \nu)(\lambda - \sigma \nu +1)/2} \,
	D^{\lambda}_{\mu \nu - \sigma} \; ; \;\;\;\;\;
	[I_x,D^{\lambda}_{\mu \nu}]=
	\nu D^{\lambda}_{\mu \nu} \label{y2}
\end{eqnarray}

It is clear that irreducible tensors in the moving frame are not
physical operators (unless $\lambda=0$), since they do not commute with
the constraining operators $J_{\nu}-I_{\nu}$ (and thus with the BRST
charge).  However.  the relevant operators to be used in the
calculation of transition probabilities are those in the laboratory
frame. Using the commutation relations (\ref{y2}) one finds that they
are physical operators, since
\begin{equation}
[(J_{\nu}-I_{\nu}),T^{\mathrm{(lab)}}_{\lambda \mu}]=0 \label{y3}
\end{equation}

We may write the operators in the laboratory frame as an expansion
\begin{eqnarray}
T^{\mathrm{(lab)}}_{\lambda\mu} &=&
		(T^{\mathrm{(lab)}}_{\lambda \mu})^{(0)} +
		(T^{\mathrm{(lab)}}_{\lambda \mu})^{(1)} +
		\cdots \nonumber \\
(T^{\mathrm{(lab)}}_{\lambda \mu})^{(0)} &=&
		(D^{\lambda}_{\mu \nu})^{(0)}
		T^{(0)}_{\lambda \nu}  \nonumber \\
(T^{\mathrm{(lab)}}_{\lambda \mu})^{(1)} &=&
		(D^{\lambda}_{\mu \nu})^{(1)}
		T^{(0)}_{\lambda  \nu} +
		(D^{\lambda}_{\mu \nu})^{(0)}
		T^{(1)}_{\lambda \nu}, \label{y4}
\end{eqnarray}
where the expressions for the matrices $(D^{\lambda}_{\mu \nu})^{(n)}$
are given in appendix B.  It is easy to verify that each separate
component $(T^{\mathrm{(lab)}}_{\lambda \mu})^{(n)}$ is not physical.
For instance, both $n=0,1$ components are needed in order to cancel the
constant term in the commutator
\begin{eqnarray}
{[(J_x-I_x),(T^{\mathrm{(lab)}}_{\lambda \mu})^{(0)}]}^{(0)} &=&
		-[I_x^{(2)},(D^{\lambda}_{\mu \nu})^{(0)}]
		T^{(0)}_{\lambda \mu} =
		-\nu (T^{\mathrm{(lab)}}_{\lambda \mu})^{(0)}
		\nonumber \\
{[(J_x-I_x),(T^{\mathrm{(lab)}}_{\lambda \mu})^{(1)}]}^{(0)} &=&
		[J_x^{(1)},T^{(1)}_{\lambda \nu}]
		(D^{\lambda}_{\mu \nu})^{(0)} =
		\nu (T^{\mathrm{(lab)}}_{\lambda \mu})^{(0)}
		\label{y5}
\end{eqnarray}

\subsection{The representation of the rotational matrices} \label{del}
States with different values of $I$  are connected through nonspherical
tensors which in the laboratory frame are expressed in terms of the
rotational matrices $D^{\lambda}_{\mu \nu}$ times intrinsic operators.
In order to represent the matrices $D^{\lambda}_{\mu \nu}$  it is
convenient to use Marshalek's generalization \cite{Ma75} of the
Holstein-Primakoff algebra (see appendix B). For $\iota \ll I$, one
obtains an operator increasing the angular momentum of the rotational
ground state $|I\rangle$, namely
\begin{equation}
E_+^{2\iota}|I\rangle = |I+\iota\rangle \label{z12}
\end{equation}
Therefore, for the spurious plus rotational sectors (cf. eq. (\ref{z11}))
\begin{eqnarray}
E^{2\iota}_+|\mathrm{sp}\rangle_I|I\rangle &=&
		|\mathrm{sp}\rangle_I|I+\iota\rangle\nonumber \\
	&=& (1- \mathrm{i} \iota \theta_x) \,
		|\mathrm{sp}\rangle_{I+\iota} \, |I+\iota\rangle
		\label{z121}
\end{eqnarray}
The transformation matrices $D^{\lambda}_{\mu \nu}$ may be expressed as
a function of $E_+^{2\iota}$ and $I_{\sigma}$ as in eq. (\ref{dmats}).

\subsection{The transitions between yrast states} \label{ssyrast}
The zero-phonon term in the transition operator in the laboratory
system is  (cf. eqs. (\ref{y4}) and (\ref{dmats}))
\begin{equation}
(T^{\mathrm{(lab)}}_{\lambda \mu})^{(0)} =
		E^{2\mu}_+ T^{(0)}_{\lambda \mu}
\label{y45}
\end{equation}
If the invariance with respect to a rotation of $\pi$ around the
$x$-axis is preserved, the components $T^{(0)}_{\lambda \mu}$ with an
odd value of $\mu$ should vanish.

For the particular case of the quadrupole operator, there are two non
vanishing components which may be associated with the static
expectation value ($T^{(0)}_{2,0}=Q_0$), and with the transition
between the yrast states $I\rightarrow I\pm 2$ ($T^{(0)}_{2,\pm
2}=Q_2$), respectively.  The magnetic dipole operator has only the
non-vanishing component corresponding to the static magnetic moment.

\subsection{The treatment of the unphysical terms in the transition
operators}  \label{ssun}
As a consequence of eqs. (\ref{y5}), we may  excite a state belonging
to the unphysical subspace (which is not annihilated by the BRST
charge) by applying for instance the operator $
(T^{\mathrm{(lab)}}_{\lambda \mu})^{(1)}$ to the ground state
$|\mathrm{sp}\rangle_I|I\rangle$.  Thus we must find out which are
these unphysical components and how the formalism eliminates their
effects.

The general expression for the operators $T^{(1)}_{\lambda \mu}$ is
\begin{equation}
T^{(1)}_{\lambda \mu} = T^{(1)}_{x\lambda \mu} +
		T^{(1)}_{\perp\lambda \mu}   \label{y61}
\end{equation}
These terms can be written as a linear combination of $(\theta_x,
J^{(1)}_x)$ and $(\theta_{\sigma},J^{(1)}_{\sigma})$, respectively,
plus terms depending on finite frequency, real modes other than the
wobbling mode. Since these last terms are straightforward to treat,
they are omitted in the following considerations. The $x$-component of
the transition operator is
\begin{eqnarray}
T^{(1)}_{x\lambda\mu} &=&
		\mathrm{i} c_{\lambda \mu} \theta_x +
		d_{\lambda\mu} J^{(1)}_x \nonumber  \\
		&=& \mathrm{i} \mu T^{(0)}_{\lambda \mu} \theta_x -
		\mathrm{i} [\theta_x,T^{(1)}_{x\lambda \mu}]^{(0)}
		J^{(1)}_x \label{y52}
\end{eqnarray}
where use has been made of eqs. (\ref{y2}) and (\ref{a122}) in order to
replace the constants $c_{\lambda \mu}$ and $d_{\lambda \mu}$. While
the second term in the r.h.s. is a null operator (cf.  eq.
(\ref{diagx})), the first term is an unphysical operator which should
be dealt with care.

The determination of the unphysical component in the perpendicular
direction is somewhat more involved.  As discussed in appendix A, it is
convenient to use the set of orthogonal pairs
$(\theta_{\sigma},f^{(1)}_{-\sigma}), (s,t)$ where $f^{(1)}_{\sigma}$
are the linear terms in the constraints (\ref{x2}) and $s,t$ are
defined in eqs.  (\ref{a6}). In terms of these variables,
$J^{(1)}_{\sigma}$ and $T^{(1)}_{\perp \lambda \mu}$  read
\begin{eqnarray}
J^{(1)}_{\sigma} &=& f_{\sigma}^{(1)} -
		\mathrm{i} \sigma I\theta_{\sigma} -
		\sqrt{\frac{I}{2}}(\mathrm{i} s+\sigma t)
		\label{y53}\\
T^{(1)}_{\perp \lambda \mu} &=&
		\mathrm{i} c_{\lambda (\mu-\sigma)} \theta_{\sigma} +
		d_{\lambda (\mu-\sigma)} J^{(1)}_{\sigma} \nonumber \\
		&=& \mathrm{i} (c_{\lambda (\mu-\sigma)} -
		\sigma I d_{\lambda(\mu-\sigma)}) \theta_{\sigma} +
		d_{\lambda (\mu-\sigma)} \left( f^{(1)}_{\sigma} -
		\sqrt{\frac{I}{2}} (\mathrm{i} s+\sigma t) \right)
		\nonumber\\
		&=& - \mathrm{i} \sigma
		\sqrt{\frac{(\lambda+\sigma  \mu)
		(\lambda-\sigma \mu +1)}{2}} \,
		T^{(0)}_{\lambda (\mu-\sigma)}
		\theta_{\sigma} \nonumber \\
		& & + \mathrm{i} [\theta_{-\sigma},
		T^{(1)}_{\perp \lambda \mu}]^{(0)}
		\left( f^{(1)}_{\sigma} - \sqrt{\frac{I}{2}}
		(\mathrm{i} s + \sigma t) \right) \label{y54}
\end{eqnarray}
where use has been made of eqs. (\ref{y2}) and (\ref{a777}).

The linear transition operators in the laboratory frame are evaluated
by combining eqs. (\ref{y4}), (\ref{y52}), (\ref{y54}) and (\ref{z131})
\begin{equation}
(T^{\mathrm{lab}}_{\lambda \mu})^{(1)} =
		E^{2\mu}_+ \left( T^{(1)}_{\lambda\mu} -
		\sqrt{\frac{(\lambda-\sigma \mu +1)
		(\lambda+\sigma \mu)}{2I^2}} \,
		T^{(0)}_{\lambda (\mu-\sigma)}I^{(1)}_{\sigma}
		\right) \label{y55}
\end{equation}
We notice that the terms proportional to $\theta_{\sigma}$ cancel out.
Therefore, the single unphysical contribution is given by (cf. eq.
(\ref{y52}))
\begin{equation}
(T^{\mathrm{lab}}_{\lambda \mu})^{(1)}_\mathrm{un} =
		\mathrm{i} \mu T^{(0)}_{\lambda \mu}
		E^{2\mu}_+ \theta_x \label{y56}
\end{equation}

Let us confine ourselves to the consideration of the spurious and
rotational subspaces
\begin{eqnarray}
(T^{\mathrm{lab}}_{\lambda \mu})^{(1)}_\mathrm{un} \,
		|\mathrm{sp}\rangle_I \, |I\rangle &=&
		\mathrm{i} \mu T^{(0)}_{\lambda \mu}
		\theta_x \, |\mathrm{sp}\rangle_I \,
		|I+\mu\rangle \nonumber \\
		&=& \mathrm{i} \mu T^{(0)}_{\lambda \mu}
		\theta_x \, |\mathrm{sp}\rangle_{I+\mu} \,
		|I+\mu\rangle \,
		(1 + {\cal O}(I^{-\frac{1}{2}})) \label{y57}
\end{eqnarray}

However, there is a compensating term due to the fact that
$(T^{\mathrm{lab}}_{\lambda \mu})^{(0)}$ is larger than
$(T^{\mathrm{lab}}_{\lambda \mu})^{(1)}$, and thus one can not neglect
the difference between the two spurious vacua when acting with the
former and keeping terms up to the same order as the leading terms of
(\ref{y57}). Using eq. (\ref{z121}), one obtains
\begin{equation}
(T^{\mathrm{lab}}_{\lambda \mu})^{(0)} \,
		|\mathrm{sp}\rangle_I \, |I\rangle =
		(1- \mathrm{i} \mu \theta_x)
		T^{(0)}_{\lambda \mu} \,
		|\mathrm{sp}\rangle_{I+\mu} \,
		|I+\mu\rangle \label{y58}
\end{equation}
Therefore the unphysical effects are cancelled out if the total
transition operator $(T^{\mathrm{lab}}_{\lambda \mu})^{(0)}$$+$$
(T^{\mathrm{lab}}_{\lambda \mu})^{(1)}$ is used.

\subsection{The excitation of the wobbling mode} \label{ssw}
The linear term  $(T^{\mathrm{lab}}_{x\lambda \mu})^{(1)}$ has either
an unphysical component or a null component (eqs. (\ref{y52}) and
(\ref{diagx})), and therefore it does not contribute to the excitation
of the  wobbling mode\footnote{It should not contribute anyhow because
of symmetry considerations.}.  The operators $f^{(1)}_{\sigma}$ in eq.
(\ref{y54}) are also null (cf.  eq. (\ref{diagyz})). Therefore the term
exciting the wobbling mode arises from $(\mathrm{i} s+\sigma t)$ (eqs.
(\ref{a6}) and (\ref{diagyz})) and from $I^{(1)}_{\sigma}$ (eq.
(\ref{z131})). The total contribution is
\begin{equation}
(T^{\mathrm{lab}}_{\lambda \mu})^{(1)}_\mathrm{w} =
		- \frac{\mathrm{i}}{2} E_+^{2 \mu}
		\left[
		\sqrt{\frac{\Delta_y}{2 \Delta_z}}
		t_{\lambda \mu (+)}^{(\mathrm{w})}
		(\Gamma_\mathrm{w}^{\dag} + \Gamma_\mathrm{w}) +
		\sqrt{\frac{\Delta_z}{2 \Delta_y}}
		t_{\lambda \mu (-)}^{(\mathrm{w})}
		(\Gamma_\mathrm{w}^{\dag} - \Gamma_\mathrm{w})
		\right] \label{y59}
\end{equation}
where
\begin{eqnarray}
t^{(\mathrm{w})}_{\lambda \mu (\pm)} &=&
		\sqrt{\frac{(\lambda - \mu + 1)(\lambda + \mu)}{I}}
		T_{\lambda (\mu-1)}^{(0)} \pm
		\sqrt{\frac{(\lambda + \mu + 1)(\lambda - \mu)}{I}}
		T_{\lambda (\mu+1)}^{(0)} \nonumber \\
		& & - \mathrm{i} \sqrt{2I} \left(
		[\theta_{-1} , T_{\lambda \mu}^{(1)}]^{(0)} \pm
		[\theta_1 , T_{\lambda \mu}^{(1)}]^{(0)} \right)
		\label{y60}
\end{eqnarray}
which vanishes for $|\mu|$=even. Therefore the excitation of a wobbling
mode from yrast states also implies the change of at least one unit of
angular momentum\footnote{This consequence of the symmetry of the
problem is also present in the previous treatments
\cite{Ma77,MJ78,Ma79,SM83}}.

For the quadrupole operator ($T^{(0)}_{2,0}=Q_0$ and $T^{(0)}_{2,\pm
2}=Q_2$, as in subsection \ref{ssyrast}),
\begin{equation}
t^{(\mathrm{w})}_{2,1(\pm)} =
		\sqrt{\frac{6}{I}} Q_0 \pm \frac{2}{\sqrt{I}} Q_2 -
		\mathrm{i} \sqrt{2I} \left(
		[\theta_{-1} , T^{(1)}_{2,1}]^{(0)} \pm
		[\theta_1 , T^{(1)}_{2,1}]^{(0)}
		\right) \label{y81}
\end{equation}
which are a factor $I^{-\frac{1}{2}}$ smaller than the yrast
transitions (\ref{y45}), as we know from the case of the rigid rotor.
For vanishing values of $I$, the commutators in eq. (\ref{y81}) vanish.
However, the new eqs. (\ref{a777}) allow in principle for the presence
of a component in $\theta_{\sigma}$ which is odd under time reversal
and therefore which may have a non-vanishing commutator with the
quadrupole terms $T_{2 \sigma}^{(1)}$.

For the excitation of the wobbling mode through the magnetic dipole
operator we have
\begin{equation}
t^{(w)}_{1,1 (\pm)}= \sqrt{\frac{2}{I}} T^{(0)}_{1,0} -
		\mathrm{i} \sqrt{2I} \left(
		[\theta_{-1}, T^{(1)}_{1,1}]^{(0)} \pm
		[\theta_1, T^{(1)}_{1,1}]^{(0)}
		\right) \label{y82}
\end{equation}
In particular, the contribution due to the intrinsic angular momentum
operator ($T^{(0)}_{1,0}=I$ and $T^{(1)}_{1,1}=J^{(1)}_1$) cancels.
There are however contributions arising from the isovector intrinsic
angular momentum operator and from spin contributions to the magnetic
operator.

\section{Conclusions} \label{scon}
The formalism developed in \cite{KBC90} for the systematic treatment of
the cranking model concerns states belonging to the same value of the
total angular momentum $I$. Not very much is added to this case in the
present contribution, but for the fact that we give explicit
expressions for the operators which are partly associated with the
spurious sector in terms of the corresponding normal modes (appendix
A).  This transformation should be needed in order to perform
perturbation theory.

In the present paper we have constructed also states corresponding to
neighboring values $I+\iota$ (for $|\iota| \ll I$). Such states are
expressed in terms of the set of states carrying the original value
$I$. As a result of this expansion, the yrast state belonging to
$I+\iota$ contains an unphysical component, if expressed in terms of
states corresponding to $I$. This component is however necessary in
order to annihilate such a state with the BRST charge associated with
the value $I+\iota$ of the collective angular momentum (subsection
\ref{iota}).

The main contribution of this paper is the systematic calculation of
the matrix elements of irreducible transition operators (section
\ref{sirr}). In the first place it is shown that the transition
operators in the laboratory frame are physical operators, unlike the
transition operators in the moving frame. Within the framework of our
perturbative approach,  we express the laboratory operators by means of
terms with different number of phonon creation and annihilation
operators. In order to accomplish this task, the rotational matrices
$D^{\lambda}_{\mu\nu}$ are written in terms of the operator
$E^{2\iota}_+ $ raising the angular momentum by $\iota$ units in the
collective subspace and of the collective components $I_{x/\sigma}$
(eqs. (\ref{isigma}), (\ref{dmats}) and (\ref{z131})).

Although the derivation of the results is somewhat involved, the final
results are fairly simple, as is generally the case with the
applications of the BRST formalism to many-body problems. Similarly to
what happens with the basic set of states, the transition operators can
be classified according to their behaviour relative to a rotation of an
angle $\pi$ around the $x$-axis. The $x$-components yield the constant
(leading order) transition amplitudes in a straightforward way
(subsection \ref{ssyrast}). The linear (first order) terms contain an
unphysical component proportional to the RPA angle $\theta_x$, which is
exactly needed in order to cancel the unphysical component preexistent
in $I+\iota$ states (subsection \ref{iota}).  The analogous unphysical
components along the perpendicular direction disappear already at the
level of the transition operator (subsection \ref{ssun}).  The
population of the real modes (other than wobbling) is straightforward
to calculate.  The components along the perpendicular direction include
also the matrix elements exciting the wobbling mode, which are
explicitly given in subsection \ref{ssw}.  They require the previous
determination of the angles $\theta_{y/z}$ which are obtained from the
generalization (\ref{a777}) of the usual RPA equations.

\appendix
\section{The normal modes of the cranking BRST hamiltonian}
\label{apcbrst}
As in all cranking calculations, it is convenient to use the symmetry
of the problem with respect to the rotation operation through an angle
$\pi$ around the $x$-axis. As a consequence, the quadratic terms may be
split into those associated with the $x$-direction and with the
$\perp$-direction.

In the first place we review the results of \cite{KBC90} concerning the
solution of the quadratic BRST cranking hamiltonian. In the second
place, we give an explicit expression of the operators associated with
the GN modes in terms of the boson operators diagonalizing the
quadratic hamiltonian. This was not done before for the $\perp$
direction. The expressions are specially useful for constructing
effective vertices and operators.

\subsection{The rotations around the $x$-axis}
In this case one can follow the same procedure as developed in
\cite{BK90} for the abelian case. The quadratic BRST hamiltonian is
divided into three terms
\begin{eqnarray}
H^{(2)}_x &=& H_{ax} + H_{bx} + H_{gx} \nonumber\\
H_{ax} &=& \frac{1}{2\Im_x}(J^{(1)}_x)^2 + \omega_{rx} \,
		(\Gamma^{\dag}_{rx}\Gamma_{rx}+\half) \nonumber\\
H_{bx} &=& -\Omega^{(1)}_x J^{(1)}_x +
		\frac{1}{F_x} \theta_xP_x -
		\frac{A_x}{2F^2_x}P_x^2\nonumber\\
H_{gx} &=& \mathrm{i} \pi_x {\bar \pi}_x +
		\frac{1}{F_x}\eta_x {\bar \eta}_x \label{a2}
\end{eqnarray}
The moment of inertia $\Im_x$ is in general different from the $\Im$
defined by eqs. (\ref{z4}). It is obtained, together with the
(linear) RPA angle $\theta_x$, through the well known RPA eqs.
\cite{MW70}
\begin{equation}
{[H_{ax},\theta_x]} = -\frac{\mathrm{i}}{\Im_x} \, J^{(1)}_x
		\; ; \;\;\;\;\;
		[\theta_x,J^{(1)}_x] = \mathrm{i} \label{a122}
\end{equation}
The zero-frequency mode appearing in the quadratic original
hamiltonian is combined with $H_{bx}$ in order to yield the normal
modes of the spurious sector. The resultant BRST quadratic hamiltonian
is given in the first of eqs. (\ref{z5}).

The original operators may be expressed in terms of the ones associated
with the normal modes, namely
\begin{eqnarray}
J_x^{(1)} &=& \sqrt{\frac{\Im_x \omega_x}{2}}
		\left(
                \Gamma_{1x} + \Gamma_{1x}^{\dag} +
                \Gamma_{0x} + \Gamma_{0x}^{\dag}
                \right) \nonumber \\
\theta_x &=& \mathrm{i} \sqrt{\frac{1}{2 \Im_x \omega_x}}
		\left(
                \Gamma_{1x} - \Gamma_{1x}^{\dag}
                \right) \nonumber \\
P_x &=& \mathrm{i} \sqrt{\frac{\Im_x}{2 \omega_x}}
		\left(
                \Gamma_{1x} - \Gamma_{1x}^{\dag} +
                \Gamma_{0x} - \Gamma_{0x}^{\dag}
                \right) \nonumber \\
\Omega_x &=& \sqrt{\frac{\omega_x}{2 \Im_x}}
		\left(
                \Gamma_{0x} + \Gamma_{0x}^{\dag}
                \right) \label{diagx}
\end{eqnarray}
and
\begin{eqnarray}
\pi_x &=& \mathrm{i}  \sqrt{\frac{\omega_x}{2}}
		\left( b_x - \bar{a}_x \right) \nonumber \\
\bar{\eta}_x &=& \sqrt{\frac{1}{2 \omega_x}}
		\left( b_x + \bar{a}_x \right) \nonumber \\
\eta_x &=& \mathrm{i} \sqrt{\frac{1}{2 \omega_x}}
		\left( a_x - \bar{b}_x \right) \nonumber \\
\bar{\pi}_x &=& \sqrt{\frac{\omega_x}{2}}
		\left( a_x + \bar{b}_x \right)
\end{eqnarray}
The spurious parameters $A_x$ and $F_x$ are related to $\omega_x$ by
\begin{equation}
\omega_x = F_x^{-1/2} \; \; ; \; \;
		A_x = \frac{\omega_x}{F_x} \label{ax}
\end{equation}

\subsection{The rotations around the $\perp$-axis}
The previous procedure must be generalized to take into account the
additional quadratic coupling terms appearing in the $\perp$-direction
\begin{eqnarray}
H^{(2)}_{\perp} &=& H_{a\perp} + H_{b\perp} + H_{g\perp}
		\nonumber \\
H_{a\perp} &=& H^{(2)} - \frac{I}{\Im}f^{(2)}_x \nonumber\\
		&=& \frac{1}{2\Im_y}
		[(f^{(1)}_y)^2 + 2\sqrt{I}t f^{(1)}_y] +
		\frac{1}{2\Im_z}
		[(f^{(1)}_z)^2+2\sqrt{I}s f^{(1)}_z ] \nonumber \\
		& & + \half \Delta^2_z I s^2 +
		\half \Delta^2_yIt^2 +
		\frac{I}{\Im} (f_y^{(1)}\theta_z-f_z^{(1)}\theta_y) +
		\frac{I}{2 \Im} \nonumber\\
H_{b\perp} &=& -\Omega^{(1)}_y f^{(1)}_y +
		\frac{1}{F_y}\theta_yP_y - \frac{A_y}{2F_y}P^2_y -
		\Omega^{(1)}_zf^{(1)}_z+\frac{1}{F_z}\theta_zP_z -
		\frac{A_z}{2F_z}P^2_z \nonumber\\
H_{g\perp} &=& \mathrm{i} \pi_y{\bar \pi}_y +
		\frac{\mathrm{i}}{F_y} \eta_y {\bar \eta}_y +
		\mathrm{i} \pi_z{\bar \pi}_z +
		\frac{\mathrm{i}}{F_z}\eta_z{\bar \eta}_z +
		\mathrm{i} \frac{I}{\Im}(\pi_y\eta_z-\pi_z\eta_y)
		\label{a5}
\end{eqnarray}
The form of this quadratic hamiltonian is determined by the choice of
the independent pairs of conjugate variables\footnote{The most
immediate choice, $(\theta_{y/z},J^{(1)}_{y/z})$ and
$(I^{(1)}_z/\sqrt{I}, I^{(1)}_y/\sqrt{I})$, is not convenient because
$J^{(1)}_y$ and $J^{(1)}_z$ do not commute.}. These are taken to be
$(\theta_{y/z},f^{(1)}_{y/z})$ and (s,t), where the $f^{(1)}_{y/z}$ are
the linear terms in the constraints (\ref{x1}) and
\begin{equation}
s \equiv \frac{1}{\sqrt{I}}I^{(1)}_z +\sqrt{I}  \theta_y
		\; ; \;\;\;\;\;
		t \equiv \frac{1}{\sqrt{I}}I^{(1)}_y -
		\sqrt{I} \theta_z \label{a6}
\end{equation}
Use is made of the Holstein-Primakoff representation for the definition
of $I^{(1)}_{y/z}$ (see appendix B).  The value of $\Delta_{y/z}$ is
given in eq. (\ref{z80}).  The (linear) angles $\theta_{y/z}$ and the
moments of inertia $\Im_{y/z}$ are obtained through the following
generalization of eq.  (\ref{a122}).
\begin{eqnarray}
{[H_{a\perp},\theta_y]} &=&
		-\frac{\mathrm{i}}{\Im_y}J^{(1)}_y +
		\mathrm{i} I\Delta^2_y \theta_z \; ; \;\;\;\;\;
		[\theta_y,J^{(1)}_y]= \mathrm{i} \nonumber \\
{[H_{a\perp},\theta_z]}&=&
		-\frac{\mathrm{i}}{\Im_z}J^{(1)}_z -
		\mathrm{i} I\Delta^2_z \theta_y \; ; \;\;\;\;\;
		[\theta_z,J^{(1)}_z]= \mathrm{i} \nonumber \\
{[\theta_y,\theta_z]}&=& 0 \label{a777}
\end{eqnarray}

The normal modes associated with the $\perp$-direction are also given
in eq. (\ref{z5}). There is also a generalization of the relations
(\ref{diagx}) expressing the original operators in terms of the normal
modes, which are labelled by the subindices $p=\pm$ ($\bar{p}=-p$)
\begin{eqnarray}
f_y^{(1)} &=& \frac{\Im h_{z p}}{\sqrt{2I}} \left(
                \Gamma_{1 p} + \Gamma_{1 p}^{\dag} +
                \Gamma_{0 p} + \Gamma_{0 p}^{\dag}
                \right) \nonumber \\
f_z^{(1)} &=& \mathrm{i} \omega_{p} \sqrt{\frac{I}{2}} \left(
                \Gamma_{1 p} - \Gamma_{1 p}^{\dag} +
                \Gamma_{0 p} - \Gamma_{0 p}^{\dag}
                \right) \nonumber \\
P_y &=& \mathrm{i} \frac{\Im h_{z p}}{\omega_{p} \sqrt{2I}} \left(
                \Gamma_{1 p} - \Gamma_{1 p}^{\dag} +
                \Gamma_{0 p} - \Gamma_{0 p}^{\dag}
                \right) \nonumber \\
P_z &=& - \sqrt{\frac{I}{2}} \left(
                \Gamma_{1 p} + \Gamma_{1 p}^{\dag} +
                \Gamma_{0 p} + \Gamma_{0 p}^{\dag}
                \right) \nonumber \\
s &=& \sqrt{\half} \left[ \sqrt{\frac{\Delta_y}{2 \Delta_z}}
                \left( \mu_{\mathrm{w} p} -
		\lambda_{\mathrm{w} p} \right) -
                \mathrm{i} \omega_{p} \right]
                \left(
                \Gamma_{1 p} - \Gamma_{1 p}^{\dag} +
                \Gamma_{0 p} - \Gamma_{0 p}^{\dag}
                \right) \nonumber \\
		& & - \sqrt{\frac{\Delta_y}{2 \Delta_z}}
		\left( \Gamma_\mathrm{w} + \Gamma_\mathrm{w}^{\dag}
		\right) \nonumber \\
t &=& \sqrt{\half} \left[ \mathrm{i}
		\sqrt{\frac{\Delta_z}{2 \Delta_y}}
                \left( \mu_{\mathrm{w} p} +
		\lambda_{\mathrm{w} p} \right) -
                \frac{\Im h_{z p}}{I} \right]
                \left(
                \Gamma_{1 p} + \Gamma_{1 p}^{\dag} +
                \Gamma_{0 p} + \Gamma_{0 p}^{\dag}
                \right) \nonumber \\
		& & + \mathrm{i}
		\sqrt{\frac{\Delta_z}{2 \Delta_y}}
		\left( \Gamma_\mathrm{w} - \Gamma_\mathrm{w}^{\dag}
		\right) \nonumber \\
\Omega_y &=& M_{y\mathrm{w}} \left( \Gamma_\mathrm{w} -
		\Gamma_\mathrm{w}^{\dag} \right) +
                M_{ymp}
                \left( \Gamma_{m p} + \Gamma_{m p}^{\dag} \right)
                \nonumber \\
\Omega_z &=& M_{z\mathrm{w}} \left( \Gamma_\mathrm{w} +
		\Gamma_\mathrm{w}^{\dag} \right) +
                M_{zmp}
                \left( \Gamma_{m p} - \Gamma_{m p}^{\dag} \right)
                \nonumber \\
\theta_y &=& N_{y\mathrm{w}} \left( \Gamma_\mathrm{w} +
		\Gamma_\mathrm{w}^{\dag} \right) +
                N_{ymp}
                \left( \Gamma_{m p} - \Gamma_{m p}^{\dag} \right)
                \nonumber \\
\theta_z &=& N_{z\mathrm{w}} \left( \Gamma_\mathrm{w} -
		\Gamma_\mathrm{w}^{\dag} \right) +
                N_{zmp}
                \left( \Gamma_{m p} + \Gamma_{m p}^{\dag} \right)
		\label{diagyz}
\end{eqnarray}
where $m=0,1$, and
\begin{eqnarray}
M_{y\mathrm{w}} &=& \frac{\sqrt{I} F_z \omega_+ \omega_-}
                {2 \Im (\omega_+^2 - \omega_-^2)}
                \left[
                \omega_+ ( \mu_{\mathrm{w}-} - \lambda_{\mathrm{w}-} ) -
                \omega_- ( \mu_{\mathrm{w}+} - \lambda_{\mathrm{w}+} )
                \right] \nonumber \\
M_{ymp} &=&   \frac{p \sqrt{I} F_z \omega_+ \omega_-}
                {2 \sqrt{2} \Im (\omega_+^2 - \omega_-^2)}
                \left[
                \omega_p ( \mu_{\bar{p}p} + \lambda_{\bar{p}p} ) -
                \omega_{\bar{p}} ( \mu_{pp} + \lambda_{pp} - 2m + 1)
                \right] \nonumber \\
M_{z\mathrm{w}} &=& \frac{\mathrm{i}}
                {2 \sqrt{I} (\omega_+^2 - \omega_-^2)}
                \left[
                h_{z+}
		( \mu_{\mathrm{w}-} + \lambda_{\mathrm{w}-} ) -
                h_{z-}
		( \mu_{\mathrm{w}+} + \lambda_{\mathrm{w}+} )
                \right] \nonumber \\
M_{zmp} &=&   \frac{\mathrm{i} p} {2 \sqrt{2I}
		(\omega_+^2 - \omega_-^2)}
                \left[
                h_{zp} ( \mu_{\bar{p}p} - \lambda_{\bar{p}p} ) -
                h_{z\bar{p}} ( \mu_{pp} - \lambda_{pp} + 2m-1)
                \right] \nonumber \\
N_{y\mathrm{w}} &=& - \sqrt{\frac{\Delta_y}{2 I \Delta_z}} +
                \frac{\mathrm{i} \sqrt{I}}
                {2 \Im (\omega_+^2 - \omega_-^2)}
                \left(
                \mu_{\mathrm{w}-} + \lambda_{\mathrm{w}-} -
		\mu_{\mathrm{w}+} - \lambda_{\mathrm{w}+}
                \right) \nonumber \\
N_{ymp} &=& \sqrt{\frac{1}{2I}}
		\left[ \sqrt{\frac{\Delta_y}{2 \Delta_z}}
                \left( \mu_{\mathrm{w}p} -
		\lambda_{\mathrm{w}p} \right) -
                \mathrm{i} \omega_p \right] +
                \nonumber \\
                & &
                \frac{\mathrm{i} p \sqrt{I}}
                {2 \sqrt{2} \Im (\omega_+^2 - \omega_-^2)}
                \left(
                \mu_{\bar{p}p} - \lambda_{\bar{p}p} -
		\mu_{pp} + \lambda_{pp} - 2m+1
                \right) \nonumber \\
N_{z\mathrm{w}} &=& - \mathrm{i}
		\sqrt{\frac{\Delta_z}{2 I \Delta_y}} \nonumber \\
		& & + \frac{F_z}
                {2 \sqrt{I} (\omega_+^2 - \omega_-^2)}
                \left[
                h_{z-} \omega_+ (\mu_{\mathrm{w}+} -
		\lambda_{\mathrm{w}+}) -
                h_{z+} \omega_- (\mu_{\mathrm{w}-} -
		\lambda_{\mathrm{w}-})
                \right] \nonumber \\
N_{zmp} &=& - \sqrt{\frac{1}{2I}}
		\left[ \mathrm{i}
		\sqrt{\frac{\Delta_z}{2 \Delta_y}}
                \left( \mu_{\mathrm{w}p} +
		\lambda_{\mathrm{w}p} \right) -
                \frac{h_{zp} \Im}{I} \right] \\
                & & - \frac{p F_z}
                {2 \sqrt{2I} (\omega_+^2 - \omega_-^2)}
                \left[
                h_{zp} \omega_{\bar{p}} (\mu_{\bar{p}p} +
		\lambda_{\bar{p}p}) -
                h_{z\bar{p}} \omega_p (\mu_{pp} +
		\lambda_{pp} - 2m +1)
                \right] \nonumber
\end{eqnarray}
For the ghost operators,
\begin{eqnarray}
\pi_y &=& - p \frac{\mathrm{i} h_{zp}}
                {\omega_{p} (\omega_+^2 - \omega_-^2)} \,
                (\bar{a}_{p} - b_{p}) \nonumber \\
\pi_z &=& \frac{\Im h_{z+} h_{z-}}
                {I (\omega_+^2 - \omega_-^2) \,
                F_y \omega_+^2 \omega_-^2 }
                p (\bar{a}_p + b_p) \nonumber \\
\eta_y &=& - \frac{\mathrm{i} \omega_{p}}{2} \,
                (a_{p} - \bar{b}_{p}) \nonumber \\
\eta_z &=& \frac{h_{y p} \Im}{2 I} \,
                (a_{p} + \bar{b}_{p}) \nonumber \\
\bar{\pi}_y &=& - \frac{1}{2 F_y}
                (\bar{b}_+ + a_+ + \bar{b}_- + a_-) \nonumber \\
\bar{\pi}_z &=& \frac{\mathrm{i} \Im
		h_{z \bar{p}} \omega_p}
                {2 I} \,
                (a_{p} - \bar{b}_{p}) \nonumber \\
\bar{\eta}_y &=& - p \frac{F_y h_{y \bar{p}}}
                {(\omega_+^2 - \omega_-^2)} \,
                (\bar{a}_{p} + b_{p}) \nonumber \\
\bar{\eta}_z &=& p \frac{\mathrm{i} F_z
		h_{z+} h_{z-} \Im}
                {I \omega_p (\omega_+^2 - \omega_-^2)}
                (\bar{a}_{p} - b_{p})
\end{eqnarray}
The coefficients $\lambda_{\mathrm{w}p}$, $\mu_{\mathrm{w}p}$,
$\lambda_{pq}$ and $\mu_{pq}$ are given in ref. \cite{KBC90} and
reproduced here for the sake of completeness
\begin{eqnarray}
\lambda_{\mathrm{w}p} &=&
		\frac{\mathrm{i}}{\omega_\mathrm{w} - \omega_p} \left(
		\frac{\Im h_{zp}}{I F_y \omega_p}
		\sqrt{\frac{\Delta_y}{2 \Delta_z}} +
		\frac{1}{F_z}
		\sqrt{\frac{\Delta_z}{2 \Delta_y}} \right)
		\nonumber \\
\mu_{\mathrm{w}p} &=&
		\frac{\mathrm{i}}{\omega_\mathrm{w} + \omega_p} \left( -
		\frac{\Im h_{zp}}{I F_y \omega_p}
		\sqrt{\frac{\Delta_y}{2 \Delta_z}} +
		\frac{1}{F_z}
		\sqrt{\frac{\Delta_z}{2 \Delta_y}} \right)
		\nonumber \\
\lambda_{pp} &=& \half (\mu_{\mathrm{w}p}^2 - \lambda_{\mathrm{w}p}^2) -
		\frac{\Im \omega_p h_{zp}}{I}
		\nonumber \\
\lambda_{p\bar{p}} &=& \frac{1}{\omega_p - \omega_{\bar{p}}} \left[
		(\omega_\mathrm{w} - \omega_p) \lambda_{\mathrm{w}p}
		\lambda_{\mathrm{w}\bar{p}} +
		(\omega_\mathrm{w} + \omega_p)
		\mu_{\mathrm{w}p} \mu_{\mathrm{w}\bar{p}}
		-\frac{\Im h_{z\bar{p}}}{I F_z}
		\right. \nonumber \\
	     & & \left.
		- \frac{\Im \omega_{\bar{p}} h_{zp}}
		{I F_y \omega_p} -
		\frac{A_y h_{zp} h_{z\bar{p}} \Im^2}{I F_y^2
		\omega_p \omega_{\bar{p}}} -
		\frac{I A_z}{F_z^2} \right]
		\nonumber \\
\mu_{pq} &=& \frac{1}{\omega_p + \omega_q} \left[
		(\omega_\mathrm{w} - \omega_p)
		\lambda_{\mathrm{w}p} \mu_{\mathrm{w}q} +
		(\omega_\mathrm{w} + \omega_p)
		\mu_{\mathrm{w}p} \lambda_{\mathrm{w}q}
		-\frac{\Im h_{zq}}{I F_z}
		\right. \nonumber \\
	     & & \left.
		+ \frac{\Im \omega_q h_{zp}}
		{I F_y \omega_p} +
		\frac{A_y h_{zp} h_{zq} \Im^2}{I F_y^2
		\omega_p \omega_q} - \frac{I A_z}{F_z^2} \right]
\end{eqnarray}
where the frequencies $\omega_p$ of the spurious sector may be
expressed in terms of the arbitrary constants $F_y,F_z$
\begin{eqnarray}
\omega_p^2 &=& \frac{1}{2} \left( \frac{I^2}{\Im^2} + \frac{1}{F_y} +
		\frac{1}{F_z} \right) \nonumber \\
		& & + p \sqrt{\frac{1}{4}
		\left( \frac{1}{F_y} - \frac{1}{F_z} \right)^2 +
		\frac{I^4}{4 \Im^4} + \frac{I^2}{2 \Im^2} \,
		\left( \frac{1}{F_y} + \frac{1}{F_z} \right) }
		\nonumber \\
h_{ip} &\equiv& F_i^{-1} - \omega_p^2 \; ; \;\;\;\;\; (i=y,z)
\end{eqnarray}
and the constants $A_i$ are related to $F_i$ and $\omega_p$ by the
equations
\begin{eqnarray}
\frac{A_z}{F_z^2} + \frac{A_y h_{zp}}{F_y^2 h_{yp}} &=&
		- \left( \frac{1}{F_y} + \frac{1}{F_z} \right)
		\frac{\Im h_{zp}}{I^2} +
		(\omega_\mathrm{w} - \omega_p)
		\frac{\lambda_{\mathrm{w}p}^2}{I} \nonumber \\
		& & + (\omega_\mathrm{w} + \omega_p)
		\frac{\mu_{\mathrm{w}p}^2}{I} \label{ayz}
\end{eqnarray}

\section{Boson expansion for the angular momentum operators and
rotational matrices}
The collective angular momentum operators satisfy the commutation
relations
\begin{eqnarray}
[ I_{\mu}' , I_{\nu}' ] &=& \mathrm{i}
		\epsilon_{\mu \nu \rho} I_{\rho}' \nonumber \\
{[} I_{\mu} , I_{\nu} ] &=& -\mathrm{i}
		\epsilon_{\mu \nu \rho} I_{\rho} \nonumber \\
{[} I_{\mu}' , I_{\nu} ] &=& 0 \label{ies}
\end{eqnarray}
where $I_{\mu}'$ are the components of the angular momentum in the
laboratory frame and $I_{\mu}$ are the components in the body fixed
frame. The basis for the collective states are the eigenstates of the
symmetric top $\vert IMK \rangle$, where $I$ is the total angular
momentum quantum number, $M$ the projection on the laboratory $z$-axis,
and $K$ the projection on the intrinsic $x$-axis.

Following Marshalek \cite{Ma75} we represent the spherical angular
momentum operators in terms of three bosons
\begin{eqnarray}
\begin{tabular}{ll}
$I_x = \frac{1}{2} a^{\dag} a - b^{\dag} b \; ;$ &
$I_z' = \frac{1}{2} a^{\dag} a - c^{\dag} c$ \\
$I_{+1} = - b^{\dag} \sqrt{\half (a^{\dag} a - b^{\dag} b)} \; ;$ &
$I_{+1}' = c^{\dag} \sqrt{\half (a^{\dag} a - c^{\dag} c)}$
\end{tabular}
\end{eqnarray}
and the states $\vert IMK \rangle$
\begin{eqnarray}
\vert IMK \rangle = \frac{(a^{\dag})^{2I} (b^{\dag})^{I-K} \,
                (c^{\dag})^{I-M}}
                {\sqrt{(2I)! (I-K)! (I-M)!}}
                \vert 0 \rangle \nonumber
\end{eqnarray}
with $0 \leq I < \infty$ and $-I \leq K,M \leq I$, and
\begin{equation}
I = \half n_a \; ; \;\;\;\;\;
K = \half n_a - n_b \; ; \;\;\;\;\;
M = \half n_a - n_c
\end{equation}
This boson realization of the angular momentum algebra is an extension
of the Holstein-Primakoff representation and is not restricted to
an $I$-subspace.

The $D_{\mu \nu}^{\lambda}$ are operators satisfying the following
properties
\begin{eqnarray}
\begin{tabular}{ll}
$[ I_{\sigma} , D_{\mu \nu}^{\lambda} ] =
                -\sigma
		\sqrt{(\lambda + \sigma \nu)
		(\lambda -\sigma \nu + 1)/2}
                D_{\mu (\nu-\sigma)}^{\lambda} \; ;$ &
${[} D_{\mu \nu}^{\lambda} , D_{\mu' \nu'}^{\lambda'} ] = 0$ \\
${[} I_x , D_{\mu \nu}^{\lambda} ] =
                \nu D_{\mu \nu}^{\lambda} \; ;$ &
$D_{\mu \nu}^{\lambda \dag} \,
                D_{\mu \nu'}^{\lambda} = \delta_{\nu \nu'}$ \\
${[} I_{\sigma}' , D_{\mu \nu}^{\lambda} ] =
		-\sigma
                \sqrt{(\lambda -\sigma \nu)
		(\lambda +\sigma \nu + 1)/2}
                D_{(\mu+\sigma)\nu}^{\lambda} \; ;$ &
$D_{\mu \nu}^{\lambda \dag} \,
                D_{\mu' \nu}^{\lambda} = \delta_{\mu \mu'}$ \\
${[} I_z' , D_{\mu \nu}^{\lambda} ] =
                \mu D_{\mu \nu}^{\lambda} \; ;$ &
$D_{\mu \nu}^{\lambda \dag} = (-)^{\nu-\mu} D_{-\mu -\nu}^{\lambda}$
\end{tabular}
\end{eqnarray}
which are reproduced using (infinite) expansions in the bosons $a$, $b$
and $c$. From the expressions for $D_{\mu \nu}^{\frac{1}{2}}$ given in
\cite{Ma75}, any $D_{\mu \nu}^{\lambda}$ with $\lambda > \frac{1}{2}$
can be found recursively using the composition law
\begin{equation}
D_{\mu \nu}^{\lambda} = \sum_{\mu_1 \mu_2} \sum_{\nu_1 \nu_2}
	(\lambda_1 \lambda_2 \mu_1 \mu_2| \lambda \mu) \,
	(\lambda_1 \lambda_2 \nu_1 \nu_2| \lambda \nu) \,
	D_{\mu_1 \nu_1}^{\lambda_1}
	D_{\mu_2 \nu_2}^{\lambda_2} \label{claw}
\end{equation}

Marshalek's bosonic representation is particularly useful when $I$ is
high and $K \sim I$. Both the angular momentum operators and the
$D_{\mu \nu}^{\lambda}$ (when $\lambda \ll I$) can be truncated to
leading orders in an expansion in terms of $b,b^{\dag}$ operators. For
the angular momentum operators we have
\begin{eqnarray}
I_x^{(0)} = I \; ; \;\;\;\;\;
I_x^{(2)} = \left( \frac{1}{2} a^{\dag} a
                - I \right) - b^{\dag} b \nonumber \\
I_{+1} \approx I_{+1}^{(1)} = - \sqrt{I} b^{\dag} \; ; \;\;\;\;\;
I_{-1} \approx I_{-1}^{(1)} = \sqrt{I} b \label{isigma}
\end{eqnarray}
To calculate $D_{\mu \nu}^{\lambda}$ we need $D_{\mu \nu}^{\frac{1}{2}}$.
Up to first order in the $b$'s,
\begin{equation}
D_{\frac{1}{2} \frac{1}{2}}^{\frac{1}{2}} \approx E_+
		\approx \frac{a^{\dag}}{\sqrt{2I}} \; ; \;\;\;\;\;
D_{\frac{1}{2} -\frac{1}{2}}^{\frac{1}{2}} \approx
		E_+ \frac{b^{\dag}}{\sqrt{2I}} \approx
		\frac{a^{\dag} b^{\dag}}{2I} \label{dmedio}
\end{equation}
where the $E_+$ operator is such that
\begin{equation}
E_+ | I \rangle = | I+ \half \rangle \; ; \;\;\;\;\;
(E_+)^{-1} | I \rangle = | I- \half \rangle
\end{equation}
Using the composition law (\ref{claw}) and eqs. (\ref{dmedio}) we
obtain the expressions for any $D_{\mu \nu}^{\lambda}$ up to first
order in the $b$'s
\begin{eqnarray}
D_{\mu \mu}^{\lambda} \approx (D_{\mu \mu}^{\lambda})^{(0)} &=&
                (E_+)^{2 \mu} \nonumber \\
D_{\mu \mu+1}^{\lambda} \approx (D_{\mu (\mu+1)}^{\lambda})^{(1)} &=&
                -(E_+)^{2 \mu} \,
                \sqrt{\frac{(\lambda+\mu+1)(\lambda-\mu)}{2 I}} b
		\nonumber \\
		&=& -(E_+)^{2 \mu} \,
                \sqrt{\frac{(\lambda+\mu+1)(\lambda-\mu)}{2 I^2}}
		I_{-1}^{(1)} \nonumber \\
D_{\mu \mu-1}^{\lambda} \approx (D_{\mu (\mu-1)}^{\lambda})^{(1)} &=&
                (E_+)^{2 \mu} \,
                \sqrt{\frac{(\lambda-\mu+1)(\lambda+\mu)}{2 I}}
                b^{\dag} \nonumber \\
		&=& - (E_+)^{2 \mu} \,
                \sqrt{\frac{(\lambda-\mu+1)(\lambda+\mu)}{2 I^2}}
		I_{+1}^{(1)} \label{dmats}
\end{eqnarray}
all other $D_{\mu \nu}^{\lambda}\approx 0$.

It is convenient to write the expression for the operators
$I^{(1)}_{\sigma}$ in terms of the (mutually orthogonal) angular and
wobbling variables. From eqs. (\ref{a6}) and (\ref{diagyz}),
\begin{eqnarray}
I^{(1)}_{\sigma} &=& -\mathrm{i}\sigma I\theta_{\sigma} -
	\sqrt{\frac{I}{2}} (\mathrm{i} s + \sigma t)
	\label{z131} \\
		 &=& -\mathrm{i}\sigma I\theta_{\sigma}\;+\;
	\frac{\mathrm{i}}{2}
	\left[
	\sqrt{\frac{I \Delta_y}{\Delta_z}} \,
	(\Gamma^{\dag}_\mathrm{w} + \Gamma_\mathrm{w}) +
	\sigma \sqrt{\frac{I \Delta_z}{\Delta_y}}
	(\Gamma^{\dag}_\mathrm{w} - \Gamma_\mathrm{w})
	\right]
	+O_\mathrm{null} \nonumber
\end{eqnarray}

\end{document}